\documentclass[showpacs,prd,aps,twocolumn]{revtex4}
\usepackage{mathrsfs,amssymb,hyperref,amsmath}

\begin{document}
\title{Stability analysis of an autonomous system in loop quantum cosmology}
\author{Kui Xiao}
 \email{87xiaokui@mail.bnu.edu.cn}
  \affiliation{Department of Physics, Beijing Normal University, Beijing 100875, China}
\author{Jian-Yang Zhu}
\thanks{Author to whom correspondence should be addressed}
 \email{zhujy@bnu.edu.cn}
  \affiliation{Department of Physics, Beijing Normal University, Beijing 100875, China}
\date{\today}

\begin{abstract}
We discuss the stability properties of an autonomous system in loop
quantum cosmology. The system is described by a self-interacting
scalar field $\phi$ with positive potential $V$, coupled with a
barotropic fluid in the Universe. With $\Gamma=VV''/V'^2$ considered
as a function of $\lambda=V'/V$, the autonomous system is extended
from three dimensions to four dimensions. We find that the dynamic
behaviors of a subset, not all, of the fixed points are independent
of the form of the potential. Considering the higher-order
derivatives of the potential, we get an infinite-dimensional
autonomous system which can describe the dynamical behavior of the
scalar field with more general potential. We find that there is just
one scalar-field-dominated scaling solution in the loop quantum
cosmology scenario.
\end{abstract}

 \maketitle

\section{Introduction}
The scalar field plays an important role in modern cosmology.
Indeed, scalar-field cosmological models are of great importance in
the study of the early Universe, especially in the investigation of
inflation. The dynamical properties of a scalar fields also make an
interesting research topic for modern cosmological studies
\cite{Copeland-IJMPD,Coley-Book}. The dynamical behavior of scalar
field coupled with a barotropic fluid in a spatially flat
Friedmann-Robertson-Walker universe has been studied by many authors
(see \cite{Copeland-IJMPD,Copeland-scalar,Leon-Phase}, and the first
section of \cite{Coley-Book}).

The phase-plane analysis of the cosmological autonomous system is an
useful method for studying the dynamical behavior of a scalar field.
One always considers the dynamical behavior of a scalar field with
an exponential potential in the classical cosmology
\cite{Copeland-exponential,Hao-PRD1,Hao-PRD2} or modified cosmology
\cite{Samart-Phantom,Li-O(N)}. And, if one considers the dynamical
behavior of a scalar field coupled with a barotropic fluid, the
exponential potential is also the first choice
\cite{Ferreira-scaling,Hoogen-scaling,Billyard-interaction,Fu-phantom}.
The exponential potential $V$ leads to the fact that the variables
$\Gamma=VV''/V'^2$ equal 1 and that $\lambda=V'/V$ is also a
constant. Then the autonomous system is always two dimensional in
classical cosmology \cite{Copeland-exponential}, and three
dimensional in loop quantum cosmology (LQC) \cite{Samart-Phantom}.
Although one can also consider a more complex case with $\lambda$
being a dynamically changing quantity
\cite{Copeland-IJMPD,Macorra-lambda,Nunes-lambda}, the fixed point
is not a real one, and this method is not exact. Recently, Zhou
\emph{et al}  \cite{Zhou-plb,Fang-CQG} introduced a new method by
which one can make $\Gamma$ a general function of $\lambda$. Then
the autonomous system is extended from two dimensions to three
dimensions in classical cosmology. They found that this method can
help investigate many quintessence models with different potentials.
The goal of this paper is to extend this method for studying the
dynamical behavior of a scalar field with a general potential
coupled with a barotropic fluid in LQC.

LQC \cite{Bojowald-Living,Ashtekar-overview} is a canonical
quantization of homogeneous spacetime based on the techniques used
in loop quantum gravity (LQG) \cite{Rovelli-Book,Thiemann-Book}.
Owing to the homogeneity and isotropy of the spacetime, the phase
space of LQC is simpler than that of LQG. For example, the
connection is determined by a single parameter $c$ and the triad is
determined by $p$. Recently, it has been shown that the loop quantum
effects can be very well described by an effective modified
Friedmann dynamics. Two corrections of the effective LQC are always
considered: the inverse volume correction and the holonomy
correction. These modifications lead to many interesting results:
the big bang can be replaced by the big bounce \cite{Ashtekar}, the
singularity can be avoided \cite{Singh}, the inflation can be more
likely to occur (e.g., see
\cite{Bojowald,Germani-inflation,Copeland-superinflation,Ashtekar-inflation,Corichi-measure}),
and more. But the inverse volume modification suffers from gauge
dependence which cannot be cured and thus yields unphysical effects.
In the effective LQC based on the holonomy modification, the
Friedmann equation adds a $-\frac{\kappa}{3}\frac{\rho^2}{\rho_c}$
term, in which $\kappa=8\pi G$, to the right-hand side of the
standard Friedmann equation \cite{Ashtekar-improve}. Since this
correction comes with a negative sign, the Hubble parameter $H$, and
then $\dot{a}$ will vanish when $\rho=\rho_c$, and the quantum
bounce occurs. Moreover, for a universe with a large scalar factor,
the inverse volume modification to the Friedmann equation can be
neglected and only the holonomy modification is important.

Based on the holonomy modification, the dynamical behavior of dark
energy has recently been investigated by many authors
\cite{Samart-Phantom,Samart-dy,Xiao-dynamical}. The attractor
behavior of the scalar field in LQC has also been studied
\cite{Copeland-superinflation,Lidsey-attractor}. It was found that
the dynamical properties of dark-energy models in LQC are
significantly different from those in classical cosmology. In this
paper, we examine the background dynamics of LQC dominated by a
scalar field with a general positive potential coupled with a
barotropic fluid. By considering $\Gamma$ as a function of
$\lambda$, we investigate scalar fields with different potentials.
Since the Friedmann equation has been modified by the quantum
effect, the dynamical system will be very different from the one in
classical cosmology, e.g., the number of dimensions of autonomous
system will change to four in LQC. It must be pointed out that this
method cannot be used to describe the dynamical behavior of scalar
field with arbitrary potential. To overcome this problem, therefore,
we should consider an infinite-dimensional autonomous system.

The paper is organized as follows. In Sec. \ref{sec2}, we present
the basic equations and the four dimensional dynamical system, and
in Sec. \ref{sec3}, we discuss the properties of this system. In
Sec. \ref{sec4}, we discuss the autonomous system in greater detail,
as well as an infinite-dimensional autonomous system. We conclude
the paper in the last section. The Appendix contains the analysis of
the dynamical properties of one of the fixed points, $P_3$.

\section{Basic equations}\label{sec2}
We focus on the flat Friedmann-Robertson-Walker cosmology. The
modified Friedmann equation in the effective LQC with holonomy
correction can be written as \cite{Ashtekar-improve}
\begin{eqnarray}
H^2=\frac{1}{3}\rho\left(1-\frac{\rho}{\rho_c}\right),\label{Fri}
\end{eqnarray}
in which $\rho$ is the total energy density and the natural unit
$\kappa=8\pi G=1$ is adopted for simplicity. We consider a
self-interacting scalar field $\phi$ with a positive potential
$V(\phi)$ coupled with a barotropic fluid. Then the total energy
density can be written as $\rho=\rho_\phi+\rho_\gamma$, with the
energy density of scalar field
$\rho_\phi=\frac12\dot{\phi}^2+V(\phi)$ and the energy density of
barotropic fluid $\rho_\gamma$. We consider that the energy momenta
of this field to be covariant conserved. Then one has
\begin{eqnarray}
&&\ddot{\phi}+3H\dot{\phi}+V'=0,\label{ddotphi}\\
&&\dot{\rho}_\gamma+3\gamma H\rho_\gamma=0, \label{dotrg}
\end{eqnarray}
where $\gamma$ is an adiabatic index and satisfies
$p_\gamma=(\gamma-1)\rho_\gamma$ with $p_\gamma$ being the pressure
of the barotropic fluid, and the prime denotesthe differentiation
with respect to the field $\phi$. Differentiating Eq. (\ref{Fri})
and using Eqs. (\ref{ddotphi}) and (\ref{dotrg}), one can obtain
\begin{eqnarray}
\dot{H}=-\frac12\left(\dot{\phi}^2+\gamma\rho_\gamma\right)\left[1-\frac{2(\rho_\gamma+\rho_\phi)}{\rho_c}\right].
\label{Fri4}
\end{eqnarray}

Equations (\ref{Fri})-(\ref{dotrg}) and (\ref{ddotphi})-(\ref{Fri4})
characterize a closed system which can determine the cosmic
behavior. To analyze the dynamical behavior of the Universe, one can
further introduce the following variables
\cite{Copeland-exponential,Samart-Phantom}:
\begin{eqnarray}
x\equiv\frac{\dot{\phi}}{\sqrt{6}H},\quad
y\equiv\frac{\sqrt{V}}{\sqrt{3}H},\quad
z\equiv\frac{\rho}{\rho_c},\quad \lambda\equiv\frac{V'}{V},
\label{new-v}
\end{eqnarray}
where the $z$ term is a special variable in LQC [see Eq.
(\ref{Fri})]. In the LQC scenario, the total energy density $\rho$
should be less than or equal to the critical energy density
$\rho_c$, and thus $0\leq z\leq 1$. Notice that, in the classical
region, $z=0$ for $\rho\ll\rho_c$. Using these new variables, one
can obtain
\begin{eqnarray}
&&\frac{\rho_\gamma}{3H^2}=\frac{1}{1-z}-x^2-y^2,\label{rg-xyz}\\
&&\frac{\dot{H}}{H^2}=-\left[3x^2+\frac{3\gamma}{2}\left(\frac{1}{1-z}-x^2-y^2\right)\right]\left(1-2z\right)\nonumber\\
\label{dH-HH}.
\end{eqnarray}

Using the new variables (\ref{new-v}), and considering Eqs.
(\ref{rg-xyz}) and (\ref{dH-HH}), one can rewrite Eqs.
(\ref{Fri})-(\ref{dotrg}) in the following forms:
\begin{eqnarray}
\frac{dx}{dN} &=&-3x-\frac{\sqrt{6}}{2}\lambda y^2+x\left[3x^2+\frac{3\gamma}{2}\left(\frac{1}{1-z}-x^2-y^2\right)\right]\nonumber\\
&&\times(1-2z),\label{x'}\\
\frac{dy}{dN}&=&\frac{\sqrt{6}}{2}\lambda x
y+y\left[3x^2+\frac{3\gamma}{2}
\left(\frac{1}{1-z}-x^2-y^2\right)\right]\nonumber\\
&&\times(1-2z),\label{y'}\\
\frac{dz}{dN} &=&-3\gamma z-3z\left(1-z\right)\left(2x^2-\gamma x^2-\gamma y^2\right),\label{z'}\\
\frac{d\lambda}{dN}  &=&\sqrt{6}\lambda^2
x\left(\Gamma-1\right),\label{l}
\end{eqnarray}
where $N=\ln a$ and
\begin{eqnarray}
  \Gamma\equiv \frac{VV''}{V'^2}.\label{Gamma}
\end{eqnarray}
Note that the potential $V(\phi)$ is positive in this paper, but one
can also discuss a negative potential. Just as \cite{Heard-negative}
has shown, the negative scalar potential could slow down the growth
of the scale factor and cause the Universe to be in a collapsing
phase. The dynamical behavior of the scalar field with the positive
and negative potential in brane cosmology has been discussed by
\cite{Copeland-scalar}. In this paper we are concerned only with an
expanding universe, and both the Hubble parameter and the potential
are positive.

Differentiating $\lambda$ with respect to the scalar field $\phi$,
we obtain the relationship between $\lambda$ and $\Gamma$,
\begin{eqnarray}
\frac{d\lambda^{-1}}{d\phi}=1-\Gamma. \label{l-G}
\end{eqnarray}
If we only consider a special case of the potential, like
exponential potential
\cite{Copeland-exponential,Hao-PRD1,Hao-PRD2,Samart-Phantom,Li-O(N),Ferreira-scaling,
Hoogen-scaling,Billyard-interaction,Fu-phantom}, then $\lambda$ and
$\Gamma$ are both constants. In this case, the four dimensional
dynamical system, Eqs. (\ref{x'})-(\ref{l}), reduces to a
3-dimensional one, since $\lambda$ is a constant. (In the classical
dynamical system, the $z$ term does not exist, and then the
dynamical system is reduced from three dimensions to two
dimensions.) The cost of this simplification is that the potential
of the field is restricted. Recently, Zhou \emph{et al}
\cite{Zhou-plb,Fang-CQG} considered the potential parameter $\Gamma$
as a function of another potential parameter $\lambda$, which
enables one to study the fixed points for a large number of
potentials. We will follow this method in this section and the
sections that follow to discuss the dynamical behavior of the scalar
field in the LQC scenario, and we have
\begin{eqnarray}
\Gamma(\lambda)=f(\lambda)+1.\label{G-l}
\end{eqnarray}
In this case, Eq. (\ref{G-l}) can cover many scalar potentials.

For completeness' sake, we briefly review the discussion of
\cite{Fang-CQG} in the following. From Eq. (\ref{l-G}), one can
obtain
\begin{eqnarray}
\frac{d\lambda}{\lambda f(\lambda)}=\frac{dV}{V}.\label{l-f-V}
\end{eqnarray}
Integrating out $\lambda=\lambda(V)$, one has the following
differential equation of potential
\begin{eqnarray}
\frac{dV}{V\lambda(V)}=d\phi.\label{l-V}
\end{eqnarray}
Then, Eqs. (\ref{l-f-V}) and (\ref{l-V}) provide a route for
obtaining the potential $V=V(\phi)$. If we consider a concrete form
of the potential (e.g., an exponential potential), the dynamical
system is specialized (e.g., the dynamical system is reduced to
three dimensions if one considers the exponential potential for
$d\lambda/dN=0$). These specialized dynamical systems are too
special if one hopes to distinguish the fixed points that are the
common properties of scalar field from those that are just related
to the special potentials \cite{Fang-CQG}. If we consider a more
general $\lambda$, then we can get the more general stability
properties of scalar field in the LQC scenario. We will continue the
discussion of this topic in Sec. \ref{sec4}. In this case, Eq.
(\ref{l}) becomes
\begin{eqnarray}
\frac{d\lambda}{dN}&=&\sqrt{6}\lambda^2 xf(\lambda).\label{l'}
\end{eqnarray}
Hereafter, Eqs. (\ref{x'})-(\ref{z'}) along with Eq. (\ref{l'})
definitely describe a dynamical system. We will discuss the
stability of this system in the following section.

\section{Properties of the autonomous system}
\label{sec3} Obviously, the terms on the right-hand side of Eqs.
(\ref{x'})-(\ref{z'}) and (\ref{l'}) only depend on $x,y,z,\lambda$,
but not on $N$ or other variables. Such a dynamical system is
usually called an autonomous system. For simplicity, we define $
\frac{dx}{dN}=F_1(x,y,z,\lambda)\equiv F_1,
\frac{dy}{dN}=F_2(x,y,z,\lambda)\equiv F_2,
\frac{dz}{dN}=F_3(x,y,z,\lambda)\equiv F_3$, and
$\frac{d\lambda}{dN}=F_4(x,y,z,\lambda)\equiv F_4.$ The fixed points
$(x_c,y_c,z_c,\lambda_c)$ satisfy $F_i=0, i=1,2,3,4$. From Eq.
(\ref{l'}), it is straightforward to see that $x=0, \lambda=0$ or
$f(\lambda)=0$ can make $F_4(x,y,z,\lambda)=0$. Also, we must
consider $\lambda^2f(\lambda)=0$. Just as \cite{Fang-CQG} argued, it
is possible that $\lambda^2 f(\lambda)\neq0$ and
$\frac{d\lambda}{dN}\neq0$ when $\lambda=0$. Thus the necessary
conditionfor the existence of the fixed points with $x\neq 0$ is
$\lambda^2f(\lambda)=0$. Taking into account these factors, we can
easily obtain all the fixed points of the autonomous system
described by Eqs. (\ref{x'})-(\ref{z'}) and (\ref{l'}), and they are
shown in Tab. \textbf{I}.

\begin{widetext}
\begin{center}
\begin{table}[h]
\caption{The stability analysis of an autonomous system in LQC. The
system is described by a self-interacting scalar field $\phi$ with
positive potential $V$ coupled with a barotropic fluid
$\rho_\gamma$. Explanation of the symbols used in this table:
$P_{i}$ denotes the fixed points located in the four dimensionsal
phase space, which are earmarked by the coordinates
$(x_c,y_c,z_c,\lambda_c)$. $\lambda_*$ means that $\lambda$ can be
any value. $\lambda_a$ is the value that makes $f(\lambda)=0$.
$\mathbf{M}^T$ means the inverted matrix of the eigenvalues of the
fixed points.
$f_1(\Lambda)=\left.\frac{df(\lambda)}{d\lambda}\right|_{\lambda=\Lambda}$
with $\Lambda=0, \lambda_a$. $A=\left[2f(\lambda_a)+\lambda_a\left(
\left.\frac{df(\lambda)}{d\lambda}\right|_{\lambda_a}\right)\right]$.
U stands for unstable, and S stands for stable.}
\begin{ruledtabular}
\begin{tabular}[c]{cccccccc}
Fixed-points & $x_c$ & $y_c$&$z_c$ &$\lambda_c$  & Eigenvalues & Stability \\
\hline
$P_1$&$0$&$0$&$0$&$0$ & $\mathbf{M}^T=(0,-3\gamma,\frac32\gamma,-3+\frac32\gamma)$& U, for all $\gamma$\\
\hline
$P_2$&$0$&$0$&$0$ &$\lambda_*$  &  $\mathbf{M}^T=(0,\frac32\gamma,-3\gamma,-3+\frac32\gamma)$& U, for all $\gamma$\\
\hline
&&&&&& S, for $\gamma=1,f_1(0)\geq0$ \\
$P_3$&$0$&$1$&0&$0$ &$\mathbf{M}^T=(-3,-3\gamma,0,0)$& U, for $\gamma=\frac43$, if $f_1(0)\neq 0$\\
&&&&&& S, for $\gamma=\frac43$, if $f_1(0)=0$\\
\hline
$P_4$&$1$&$0$&$0$&$0$ &$\mathbf{M}^T=(0,-6,0,6-3\gamma)$&U,for all $\gamma$\\
\hline
$P_5$&$-1$&$0$&$0$&$0$ &$\mathbf{M}^T=(0,-6,0,6-3\gamma)$&U,for all $\gamma$\\
\hline
$P_6$&$0$&$0$&$0$&$\lambda_a$  &$\mathbf{M}^T=(0,\frac32\gamma,-3\gamma,-3+\frac32\gamma)$&U,for all $\gamma$\\
\hline $P_7$&$1$&$0$&$0$&$\lambda_a$ &$\mathbf{M}^T=
\left(-6,6-3\gamma,\frac12\sqrt{6}\lambda_a+3,\sqrt{6}\lambda_a A\right)$&U,for all $\gamma$\\
\hline $P_8$&$-1$&$0$&$0$&$\lambda_a$&$\mathbf{M}^T=
\left(-6,6-3\gamma,-\frac12\sqrt{6}\lambda_a+3,-\sqrt{6}\lambda_aA\right)$&U,for all $\gamma$\\
\hline
$P_9$&$-\frac{\sqrt{6}}{6}\lambda_a$ &$\sqrt{1-\frac{\lambda^2_a}{6}}$&0&$\lambda_a$ &$\mathbf{M}^T=\left(-\lambda_a^2,-3+\frac12\lambda_a^2,\lambda_a^2-3\gamma,-\lambda^3_a-f_1(\lambda_a) \right)$& S, for $f_1(\lambda_a)>\lambda_a$, and $\lambda_a<3\gamma$ \\
&&&&&& U, for $f_1(\lambda_a)<\lambda_a$ and/or $\lambda_a>3\gamma$\\
\hline $P_{10}$ &$-\sqrt{\frac32}\frac{\gamma}{\lambda_a}$
&$\sqrt{\frac{3}{2\lambda_a^2}\gamma(2-\gamma)}$&0&$\lambda_a$& See
the Eq. (\ref{p10}) & U, for all $\gamma$
\end{tabular}
\end{ruledtabular}
\end{table}
\end{center}
\end{widetext}

The properties of each fixed points are determined by the
eigenvalues of the Jacobi matrix
\begin{eqnarray}
\mathcal{M}= \left. \begin{pmatrix}
\frac{\partial F_1}{\partial x}&\frac{\partial F_1}{\partial y}&\frac{\partial F_1}{\partial z}&\frac{\partial F_1}{\partial \lambda}\\
\frac{\partial F_2}{\partial x}&\frac{\partial F_2}{\partial y}&\frac{\partial F_2}{\partial z}&\frac{\partial F_2}{\partial \lambda}\\
\frac{\partial F_3}{\partial x}&\frac{\partial F_3}{\partial y}&\frac{\partial F_3}{\partial z}&\frac{\partial F_3}{\partial \lambda}\\
\frac{\partial F_4}{\partial x}&\frac{\partial F_4}{\partial
y}&\frac{\partial F_4}{\partial z}&\frac{\partial F_4}{\partial
\lambda}
\end{pmatrix}
\right|_{(x_c,y_c,z_c,\lambda_c)}.
\end{eqnarray}
According to Lyapunov's linearization method, the stability of a
linearized system is determined by the eigenvalues of the matrix
$\mathcal{M}$ (see Chapter 3 of \cite{Slotine-book}). If all of the
eigenvalues are strictly in the left-half complex plane, then the
autonomous system is stable. If at least one eigenvalue is strictly
in the right-half complex plane, then the system is unstable. If all
of the eigenvalues are in the left-half complex plane, but at least
one of them is on the $i\omega$ axis, then one cannot conclude
anything definite about the stability from the linear approximation.
By examining the eigenvalues of the matrix $\mathcal{M}$ for each
fixed point shown in Table \textbf{I}, we find that points
$P_{1,2,4-8,10}$ are unstable and point $P_{9}$ is stable only under
some conditions. We cannot determine the stability properties of
$P_{3}$ from the eigenvalues, and we will give the full analysis of
$P_3$ in the Appendix.

Some remarks on Tab.\textbf{I}:
\begin{enumerate}
\item
Apparently, points $P_2$ and $P_6$ have the same eigenvalues, and
the difference between these two points is just on the value of
$\lambda$.  As noted in the caption of Table \textbf{I}, $\lambda_*$
means that $\lambda$ can be any value, and $\lambda_a$ is just the
value that makes $f(\lambda)=0$. Obviously, $\lambda_a$ is just a
special value of $\lambda_*$, and point $P_6$ is a special case of
point $P_2$. $P_6$ is connected with $f(\lambda)$, but $P_2$ is not.
From now on,  we do not consider separately the special case of
$P_6$ when we discuss the property of $P_2$. Hence the value of
$\lambda_a$ is contained in our discussion of$\lambda_*$.
\item
It is straightforward to check that, if $x_c=\lambda_c=0$, $y_c$ can
be any value $y_*$ when it is greater than or equal $1$. But, if
$y_*>1$, then $z_c=1-1/y_*^2<1$, and this means that the fixed point
is located in the quantum-dominated regions. Although the stability
of this point in the quantum regions may depend on $f(\lambda)$, it
is not necessary to analyze its dynamical properties, since it does
not have any physical meanings. The reason is the following: If the
Universe is stable, it will not evolve to today's pictures. If the
Universe is unstable, it will always be unstable. We will just focus
on point $P_3$ staying in the classical regions. Then
$y_c=y_*=1,z_c=1-1/y_*^2=0$, i.e., for the classical cosmology
region, $\rho\ll\rho_c$.
\item
Since the adiabatic index $\gamma$ satisfies $0<\gamma< 2$ (in
particular, for radiation $\gamma=\frac43$ and for dust $\gamma=1$),
all the terms that contain $\gamma$ should not change sign. A more
general situation of $\gamma$ is $0\leq \gamma\leq 2$
\cite{Billyard-scaling}. If $\gamma=0$ or $\gamma=2$, the
eigenvalues corresponding to points $P_{1,2,4,5,9}$ will have some
zero elements and some negative ones. To analyze the stability of
these points, we need to resort to other more complex methods, just
as we do in the Appendix for the dynamical properties of point
$P_3$. In this paper, we just consider the barotropic fluid which
includes radiation and dust, and $\gamma\neq 0,2$. Notice that if
one considers $\gamma=0$, the barotropic fluid describes the dark
energy. This is an interesting topic, but will not be considered
here for the sake of simplicity.
\item
$-\sqrt{6}<\lambda_a<\sqrt{6},\lambda_a\neq 0$ should hold for
point $P_9$, hence $-3+\frac12\lambda^2_a<0$.
\item
$\lambda_a>0$ should hold, since $y_c>0$ for point $P_{10}$. The
eigenvalue of this point is
\begin{eqnarray}\label{p10}
\mathbf{M}=\begin{pmatrix}
-3\gamma\\-3\lambda_a\gamma f_1(\lambda_a)\\-\frac32+\frac34\gamma+\frac{3}{4\lambda_a}\sqrt{(2-\gamma)(\lambda^2_a(2-\gamma)+8\gamma+24\gamma^2)}\\
-\frac32+\frac34\gamma-\frac{3}{4\lambda_a}\sqrt{(2-\gamma)(\lambda^2_a(2-\gamma)+8\gamma+24\gamma^2)}
\end{pmatrix}.\nonumber\\
\end{eqnarray}
Since we just consider $0<\gamma<2$ in this paper, it is easy to
check that $(2-\gamma)(\lambda^2_a(2-\gamma)+8\gamma+24\gamma^2)>0$
is always satisfied. And this point is unstable with
$f_1(\lambda_a)=\left.\frac{df(\lambda)}{d\lambda}\right|_{\lambda=\lambda_a}$
being either negative or positive, since
$-\frac32+\frac34\gamma+\frac{3}{4\lambda_a}\sqrt{(2-\gamma)(\lambda^2_a(2-\gamma)+8\gamma+24\gamma^2)}$
is always positive.
\end{enumerate}

Based on Table \textbf{I} and the related remarks above, we have the
folloing conclusions:
\begin{enumerate}
\item Points $P_{1,2}$: The related critical values,
eigenvalues and stability properties do not depend on the specific
form of the potential, since $\lambda_c=0$ or $\lambda$ can be any
value $\lambda_*$.
\item Point $P_3$: The related stability properties depend on
$f_1(0)=\left.\frac{df(\lambda)}{d\lambda}\right|_{\lambda=0}$.
\item Points $P_{4,5}$: The related eigenvalues and stability properties
do not depend on the form of the potential, but the critical values
of these points should satisfy $\lambda^2 f(\lambda)=0$ since
$x_c\neq 0$.
\item Point $P_6$: It is a special case of $P_2$, but $f(\lambda_a)=0$ should be satisfied.
\item Points $P_{7,8}$: Same as $P_6$, they would not exist if $f(\lambda_a)\neq 0$.
\item Point $P_{9,10}$: $f(\lambda_a)=0$ should hold. The fixed values and the eigenvalues of these two points depend on $f_1(\lambda_a)=
\left.\frac{df(\lambda)}{d\lambda}\right|_{\lambda=\lambda_a}$.
\end{enumerate}
Thus, only points $P_{1,2}$ are independent of $f(\lambda)$.

Comparing the fixed points in LQC and the ones in classical
cosmology (see the Table \textbf{I} of \cite{Fang-CQG}), we can see
that, even though the values of the coordinates
$(x_{c},y_{c},\lambda_{c})$ are the same, the stability properties
are very different. This is reasonable, because the quantum
modification is considered, and the autonomous system in the LQC
scenario is very different from the one in the classical scenario,
e.g., the autonomous system is four dimensional in LQC but three
dimensional in the classical scenario. Notice that all of the fixed
points lie in the classical regions, and therefore the coordinates
of fixed points remain the same from classical to LQC, which we also
pointed out in an earlier paper \cite{Xiao-dynamical}.

Now we focus on the late time attractors: point $P_{3}$ under the
conditions of $\gamma=1, f_1(0)\geq 0$ and $\gamma=4/3, f_1(0)=0$,
and point $P_{9}$ under the conditions of $\lambda_a^2<6,
f_1(\lambda_a)>\lambda_a,\lambda_a<3\gamma$. Obviously, these points
are scalar-field dominated, since
${\rho_\gamma}=H^2(1/(1-z_c)-x_c^2-y_c^2)=0$. For point $P_{3}$, the
effective adiabatic index
$\gamma_\phi=(\rho_\phi+p_\phi)/\rho_\phi=0$, which means that the
scalar field is an effective cosmological constant. For point
$P_{9}$, $\gamma_\phi=\lambda^2_a/2$. This describes a scaling
solution that, as the universe evolves, the kinetic energy and the
potential energy of the scalar field scale together. And we can see
that there is not any barotropic fluid coupled with the
scalar-field-dominated scaling solution. This is different from the
dynamical behavior of scalar field with exponential potential $V=V_0
\exp(-\lambda\kappa\phi)$ in classical cosmology
\cite{Copeland-exponential,Hao-PRD1,Hao-PRD2,Samart-Phantom,Li-O(N),Ferreira-scaling,
Hoogen-scaling,Billyard-interaction,Fu-phantom}, and also is
different from the properties of the scalar field in brane cosmology
\cite{Copeland-scalar}, in which $\lambda=\rm{const.}$ (notice that
the definition of $\lambda$ in \cite{Copeland-scalar} is different
from the one in this paper) and $\Gamma$ is a function of
$L(\rho(a))$ and $|V|$. In these models, the Universe may enter a
stage dominated by a scalar field coupled with fluid when
$\lambda,\gamma$ satisfy some conditions
\cite{Copeland-exponential,Copeland-scalar}.

We discuss the dynamical behavior of the scalar field by considering
$\Gamma$ as a function of $\lambda$ in this and the preceding
sections. But $\Gamma$ can not always be treated as a function of
$\lambda$. We need to consider a more general autonomous system,
which we will introduce in the next section.

\section{Further discussion of the autonomous system}\label{sec4}
The dynamical behavior of the scalar field has been discussed by
many authors (e.g., see \cite{Copeland-IJMPD,
Coley-Book,Copeland-exponential,Hao-PRD1,Hao-PRD2,
Samart-Phantom,Li-O(N),Ferreira-scaling,Hoogen-scaling,Billyard-interaction,Fu-phantom}).
If one wants to get the potentials that yield the cosmological
scaling solutions beyond the exponential potential, one can add a
$\frac{d\phi}{dN}$ term into the autonomous system
\cite{Nunes-scaling}. All of these methods deal with special cases
of the dynamical behavior of scalar fields in backgrounds of some
specific forms. By considering $\Gamma$ as a function of $\lambda$,
one can treat potentials of more general forms and get the common
fixed points of the general potential, as shown in
\cite{Zhou-plb,Fang-CQG} and in the two preceding sections. However,
as is discussed in \cite{Fang-CQG}, sometimes $\Gamma$ is not a
function of $\lambda$, and then the dynamical behaviors of the
scalar fields discussed above are still not general in the strict
sense. For a more general discussion, we must consider the
higher-order derivatives of the potential. We define
\begin{eqnarray}
&&{}^{(1)}\Gamma=\frac{VV_3}{V'^2}, \quad {}^{(2)}\Gamma
=\frac{VV_4}{V'^2},\quad{}^{(3)}\Gamma=\frac{VV_5}{V'^2},\nonumber\\
&&\cdots\quad{}^{(n)}\Gamma=\frac{VV_{n+2}}{V'^2},\quad \cdots
\end{eqnarray}
in which $V_{n}=\frac{d^n V}{d\phi^n},n=3,4,5,\cdots$. Then we can
get
\begin{eqnarray}
\frac{d\Gamma}{d N}
&=&\sqrt{6}x\left[\Gamma\lambda+{}^{(1)}\Gamma-2\lambda\Gamma^2\right],\label{G'}\\
\frac{d\left({}^{(1)}\Gamma\right)}{dN}
&=&\sqrt{6}x\left[{}^{(1)}\Gamma\lambda +{}^{(2)}\Gamma
-2\lambda\Gamma\left({}^{(1)}\Gamma\right)\right],\label{G3'}\\
\frac{d\left({}^{(2)}\Gamma\right)}{dN} &=&\sqrt{6}x\left[
{}^{(2)}\Gamma\lambda+{}^{(3)}\Gamma
-2 \lambda\Gamma\left({}^{(2)}\Gamma\right)\right],\label{G4'}\\
\frac{d\left({}^{(3)}\Gamma\right)}{dN}
&=&\sqrt{6}x\left[{}^{(3)}\Gamma\lambda+{}^{(4)}\Gamma
-2\lambda\Gamma\left({}^{(3)}\Gamma\right)\right],\label{G5'}\\
&&  \cdots\cdots\nonumber\\
\frac{d\left({}^{(n)}\Gamma\right)}{dN}
&=&\sqrt{6}x\left[{}^{(n)}\Gamma\lambda+{}^{(n+1)}\Gamma
-2\lambda\Gamma \left({}^{(n)}\Gamma\right)\right],\label{Gn'}\\
&& \cdots\cdots\nonumber
\end{eqnarray}
To discuss the dynamical behavior of scalar field with more general
potential, e.g., when neither $\lambda$ nor $\Gamma$ is constant, we
need to consider a dynamical system described by Eqs.
(\ref{x'})-(\ref{l}) coupled with Eqs. (\ref{G'})-(\ref{Gn'}). It is
easy to see that this dynamical system is also an autonomous one. We
can discuss the values of the fixed points of this autonomous
system. Considering Eq. (\ref{l}), we can see that the values of
fixed points should satisfy $x_c=0$, $\lambda_c=0$, or $\Gamma_c=1$.
Then, we can get the fixed points of this infinite-dimensional
autonomous system.
\begin{enumerate}
\item If $x_c=0$, considering Eqs. (\ref{x'})-(\ref{z'}), one can get
$(y_c,z_c,\lambda_c)=(0,0,0)$ or
$(y_c,z_c,\lambda_c)=(0,0,\lambda_*)$, and
$\Gamma_c,{}^{(n)}\Gamma_c$ can be any values.
\item If $\lambda_c=0$, considering Eqs. (\ref{x'})-(\ref{z'}),
one can see that the fixed points of $(x,y,z)$ are
$(x_c,y_c,z_c)=(0,y_*,1-1/y_*^2)$and $(x_c,y_c,z_c)=(\pm1,0,0)$. If
$x_c=0$, $\Gamma_c$ and ${}^{(n)}\Gamma_c$ can be any values, and if
$x_c=\pm 1$, ${}^{(n)}\Gamma_c=0$.
\item If $\Gamma_c=1$, considering Eqs. (\ref{x'})-(\ref{z'}),
one can get that the fixed points of $(x,y,z,\lambda)$ are
$(x_c,y_c,z_c,\lambda_c)=(0,0,0,\lambda_*)$ and
$(x_c,y_c,z_c,\lambda_c)=(\pm 1,0,0,\lambda_*)$. And
${}^{(n)}\Gamma_c$ should satisfy ${}^{(n)}\Gamma_c=\lambda_*^n$.
There are other fixed points, which will be discussed below.
\end{enumerate}

Based on the above analysis and Table \textbf{I}, one can find that
points $P_{1-10}$ are just special cases of the fixed points of an
infinite-dimensional autonomous systems. Considering the definition
of $\Gamma$ (see Eq. (\ref{Gamma})), the simplest potential is an
exponential potential when $\Gamma_c=1$. The properties of these
fixed points have been discussed by many authors
\cite{Copeland-exponential,Hao-PRD1,Hao-PRD2,Samart-Phantom,Li-O(N),Ferreira-scaling,Hoogen-scaling,Billyard-interaction,Fu-phantom}.
If $x_c=0$ and $y_c=0$, this corresponds to a fluid-dominated
universe, which we do not consider here. If $x_c=\pm 1$,
$\Gamma_c=0$ and ${}^{(n)}\Gamma_c=0$, we do not need to consider
the $\Gamma$ and the $^{(n)}\Gamma$ terms. Then the stability
properties of these points are the same as points $P_{4,5}$ in Table
\textbf{I}, and there are unstable points. The last case is
$(x_c,y_c,z_c,\lambda_c)=(0,y_*,1-1/y_*^2,0)$ and
$\Gamma,{}^{(n)}\Gamma$ can be any value. To analyze the dynamical
properties of this autonomous system, we need to consider the
${}^{(n)}\Gamma_c$ terms. We will get an infinite series. In order
to solve this infinite series, we must truncate it by setting a
sufficiently high-order ${}^{(M)}\Gamma$ to be a constant, for a
positive integer $M$, so that $d\left({}^{(M)}\Gamma\right)/dN=0$.
Thus we can get an $(M+4)$-dimensional autonomous system. One
example is the quadratic potential $V=\frac12m^2\phi^2$ with some
positive constant $m$ that gives a five dimensional autonomous
system, and another example is the Polynomial (concave) potential
$V=M^{4-n}\phi^n$ \cite{Lindle-potential} that gives an
$(n+3)$-dimensional autonomous system. Following the method we used
in the two preceding sections, we can get the dynamical behavior of
such finite-dimensional systems.

In the remainder of this section, we discuss whether this autonomous
system has a scaling solution.

If $x_c=0$, then $\Gamma_c\neq 0,{}^{(n)}\Gamma_c\neq 0$, and the
stability of the fixed points may depend on the truncation. As an
example, if we choose ${}^{(2)}\Gamma=0$, then we can get a six
dimensional autonomous system. The eigenvalues for the fixed point
$(x_c,y_c,z_c,\lambda_c,\Gamma_c,{}^{(1)}\Gamma_c)=(0,0,0,\lambda_b,\Gamma_*,{}^{(1)}\Gamma_*)$,
where $\lambda_b=0$ or $\lambda_b=\lambda_*$, is
\begin{eqnarray}
\mathbf{M}^T=(0,0,0,\frac32\gamma,-3\gamma,-3+\frac32\gamma).\nonumber
\end{eqnarray}
Obviously, this is an unstable point, and it has no scaling
solution. The eigenvalues for the fixed point
$(x_c,y_c,z_c,\lambda_c,\Gamma_c,{}^{(1)}\Gamma_c)=(0,1,0,0,\Gamma_*,{}^{(1)}\Gamma_*)$
is
\begin{eqnarray}
\mathbf{M}^T=(0,0,0,0,-3\gamma,-3-3\gamma).\nonumber
\end{eqnarray}
According to the center manifold theorem (see Chapter 8 of
\cite{Khalil-non}, or \cite{DynamicalReduction}), there are two
nonzero eigenvalues, and we need to reduce the dynamical system to
two dimensions to get the stability properties of the autonomous
system. This point may have scaling solution, but we need more
complex mathematical method. We discuss this problem in another
paper \cite{Xiao-scaling}.

We discuss the last case. If $\Gamma_c=1$, we can consider an
exponential potential. Then the autonomous system is reduced to
three dimensions. It is easy to check that the values
$(x_{ec},y_{ec},z_{ec})$ of the fixed points are just the values
$(x_c,y_c,z_c)$ of points $P_{6-10}$ in Table \textbf{I}. We focus
on the two special fixed points:
\begin{eqnarray}
&& F_1: (x_{ec},y_{ec},z_{ec})=(-\lambda/\sqrt{6}, \sqrt{1-\lambda^2/6}, 0),\nonumber\\
&&F_2: (x_{ec},y_{ec},z_{ec})=(-\sqrt{3/2}\gamma/\lambda,
\sqrt{3\gamma(2-\gamma)/(2\lambda^2)}, 0).\nonumber
\end{eqnarray}
Using Lyapunov's linearization method, we can find that $F_2$ is
unstable and $F_1$ is stable if $\lambda<3\gamma$. It is easy to
check that $\rho_\gamma=H^2[1/(1-z_{ec})-x_{ec}^2-y_{ec}^2]=0$ when
$(x_{ec},y_{ec},z_{ec})=(-\lambda/\sqrt{6}, \sqrt{1-\lambda^2/6},
0)$. From the above analysis, we find that there is just the
scalar-field-dominated scaling solution when we consider the
autonomous system to be described by a self-interacting scalar field
coupled with a barotropic fluid in the LQC scenario.

\section{Conclusions}\label{sec5}
The aim of this paper is two-fold. We discuss the dynamical behavior
of scalar field in the LQC scenario following \cite{Fang-CQG,
Zhou-plb}.  To further analyze the dynamical properties of scalar
field with more general potential, we introduce an
infinite-dimensional autonomous system.

To discuss the dynamical properties of scalar field in the LQC
scenario, we take $\Gamma$ as a function of $\lambda$, and extend
the autonomous system from three dimensions to four dimensions. We
find this extended autonomous system has more fixed points than the
three dimensional one does. And we find that for some fixed points,
the function $f(\lambda)$ affects either their values, e.g., for
points $P_{4-10}$, or their stability properties, e.g., for points
$P_{3,9}$. In other words, the dynamical properties of these points
depend on the specific form of the potential. But some other fixed
points, e.g., points $P_{1,2}$,are independent of the potential. The
properties of these fixed points are satisfied by all scalar fields.
We also find that there are two later time attractors, but the
Universe is scalar-field dominated since $\rho_\gamma=0$ at these
later time attractors.

The method developed by \cite{ Fang-CQG,Zhou-plb} can describe the
dynamical behavior of the scalar field with potential of a more
general form than, for example, an exponential potential
\cite{Copeland-exponential,Hao-PRD1,Hao-PRD2,Samart-Phantom,Li-O(N),Ferreira-scaling,Hoogen-scaling,Billyard-interaction,Fu-phantom}.
But it is not all-encompassing. If one wants to discuss the
dynamical properties of a scalar field with an arbitrary potential,
one needs to consider the higher-order derivatives of the potential
$V(\phi)$. Hence the dynamical system will extend from four
dimensions to infinite-dimensions. This infinite-dimensional
dynamical system is still autonomic, but it is impossible to get all
of its dynamical behavior unless one considers $\Gamma_c=1$ which
just gives an exponential potential. If one wants to study as much
as possible the dynamical properties of this infinite-dimensional
autonomous system, one has to consider a truncation that sets
${}^{(M)}\Gamma=\rm{Const.}$, with $M$ above a certain positive
integer. Then the infinite-dimensional system can be reduced to
$M+4)$ dimensions. And we find that there is just the
scalar-field-dominated scaling solution for this autonomous system.
We only give out the basic properties of this infinite-dimensional
autonomous system in this paper, and will continue the discussion in
the paper in \cite{Xiao-scaling}.

We only get the scalar-field-dominated scaling solutions, whether we
consider $\Gamma$ as a function of $\lambda$ or consider the
higher-order derivatives of the potential. This conclusion is very
different from the autonomous system which is just described by a
scalar field with an exponential potential \cite{Samart-Phantom}.
This is an interesting problem that awaits further analysis.

\begin{acknowledgments}
K. Xiao thanks Professor X. Li for his help with the center manifold
theorem. This work was supported by the National Natural Science
Foundation of China, under Grant No. 10875012 and the Fundamental
Research Funds for the Central Universities.
\end{acknowledgments}

\appendix*
\section{The stability properties of the Point $P_3$}
\setcounter{equation}{0}
\renewcommand{\theequation}{A\arabic{equation}}
In Sec. \ref{sec3}, we point out that it is impossible to get the
stability properties of the fixed point if at least one of the
eigenvalues of $\mathcal{M}$ is on the $i\omega$ axis with the rest
being in the left-half complex plane. The fixed point $P_3$ is
exactly such a point. In this appendix, we use the center manifold
theorem (see Chapter 8 of \cite{Khalil-non} , or
\cite{DynamicalReduction}) to get the condition for stability of
$P_3$. The coordinates of $P_3$ are $(0,1,0,0)$ and the eigenvalues
are $(-3,-3\gamma,0,0)$. First, we transfer $P_3$ to $P'_3$
$(x_c=0,\bar{y}_c=y-1=0,z_c=0, \lambda_c=0)$. In this case, Eqs.
(\ref{x'})-(\ref{z'}) and (\ref{l'}) become
\begin{eqnarray}
\frac{dx}{dN}&=&-3\,x-\frac12\,\sqrt {6}\lambda\, \left( \bar{y}+1
\right) ^{2}+x \left[ 3\,{x}^
{2}+\frac32\gamma\, \left((1+z)\right.\right.\nonumber\\
&&\left.\left.-{x}^{2}-\left( \bar{y}+1 \right) ^{2} \right)  \right]\left( 1-2z\right)\label{X'},\\
\frac{d \bar{y}}{dN}&=&\frac12\,\sqrt {6}\lambda\,x \left( \bar{y}+1
\right) + \left( \bar{y}+1 \right)
\left[3\,{x}^{2}+\frac32\gamma\, \left((1+z)\right.\right.\nonumber\\
&&\left.\left.-{x}^{2}-\left( \bar{y}+1 \right) ^{2} \right)
\right]  \left( 1-2z
\right)\label{Y'}, \\
\frac{dz}{dN}&=&-3\gamma z-3z\left(1-z\right)\left[2x^2-\gamma
x^2-\gamma
(\bar{y}+1)^2\right],\label{Z'}\\
\frac{d\lambda}{dN}&=& \sqrt
{6}{\lambda}^{2}\left(f(0)+f_1(0)\lambda\right) x \label{L'},
\end{eqnarray}
where we have considered that the related variables
$(x,\bar{y},z,\lambda)$ are small around point
$(x_c,\bar{y}_c,z_c,\lambda_c)=\left(0,0, 0,0\right)$. Therefore the
following Taylor series
\begin{eqnarray}
&&\frac{1}{1-z}=1+{z}+\cdots,\nonumber\\
&&f(\lambda)=f(0)+f_1(0)\lambda+\cdots,\nonumber
\end{eqnarray}
can be used, where
$f_1(0)=\frac{df(\lambda)}{d\lambda}\left.\right|_{\lambda=0}$.

We can get the Jacobi matrix $\mathcal{M'}$ of the dynamical system
Eqs. (\ref{X'})-(\ref{L'}) as
\begin{eqnarray}
\mathcal{M'}= \begin{pmatrix}
-3& 0& 0&-\frac{\sqrt{6}} 2\\
0&-3\gamma& \frac32\gamma & 0\\
0&0& 0&0\\
0&0&0&0
\end{pmatrix}.
\end{eqnarray}
It is easy to get the eigenvalues and eigenvectors of
$\mathcal{M}'$. Let $\mathcal{A}$ denote the matrix whose columns
are the eigenvalues, and $\mathcal{S}$ denote the matrix whose
columns are the eigenvectors, and then we have
\begin{eqnarray}
\mathcal{A}=\begin{pmatrix} -3\\ -3\gamma\\ 0\\ 0
\end{pmatrix},\quad
\mathcal{S}=\begin{pmatrix}
1& 0& -\frac{\sqrt{6}}{6}& 0\\
0& 1& 0 & \frac{1}{2}\\
0 &  0&0& 1\\
0 & 0& 1 & 0
\end{pmatrix}.
\end{eqnarray}
With the help of $\mathcal{S}$, we can transform $\mathcal{M}'$ into
a block diagonal matrix
\begin{eqnarray}
\mathcal{S}^{-1} \mathcal{M}' \mathcal{S}=\begin{pmatrix}
-3 & 0&0&0\\
0& -3\gamma& 0& 0\\
0&0&0&0\\
0&0&0&0
\end{pmatrix}=\begin{pmatrix}
\mathcal{A}_1 &0\\
0& \mathcal{A}_2
\end{pmatrix},
\end{eqnarray}
where all eigenvalues of $\mathcal{A}_1$ have negative real parts,
and all eigenvalues of $\mathcal{A}_2$ have zero real parts.

Now we change the variables to be
\begin{eqnarray}
    \begin{pmatrix}
    X\\Y\\Z\\ \bar{\lambda}
  \end{pmatrix}=
  \mathcal{S}^{-1}\begin{pmatrix}
    x\\ \bar{y}\\ {z}\\ \lambda
  \end{pmatrix}=
  \begin{pmatrix}
    x+\frac{\sqrt{6}}{6}\lambda\\ \bar{y}-\frac12z\\ \lambda \\z
  \end{pmatrix}.
\end{eqnarray}
Then, we can rewrite the autonomous system in the form of the new
variables:
\begin{eqnarray}
\begin{pmatrix}\label{A9}
  \frac{dX}{dN}\\ \frac{dY}{dN}\\ \frac{dZ}{dN}\\ \frac{d\bar{\lambda}}{dN}
\end{pmatrix}=
\begin{pmatrix}
  -3 & 0&0&\\
0&-3\gamma&0&0\\
  0&0&0&0\\
  0&0&0&0
\end{pmatrix}
\begin{pmatrix}
  X\\Y\\Z\\ \bar{\lambda}
\end{pmatrix}+
\begin{pmatrix}
  G_1\\G_2\\G_3\\G_4
\end{pmatrix},\nonumber\\
\end{eqnarray}
where $G_i=G_i(X,Y,Z,\bar{\lambda}),(i=1,2,3,4)$ are functions of
$X,Y,Z$, and $\bar{\lambda}$. It is easy to get $G_i$ by
substituting the transformations $x=X-\frac{\sqrt{6}}{6}Z,
\bar{y}=Y+\frac12\bar{\lambda}, z=\bar{\lambda}, \lambda=Z$ into the
R.H.S. of Eqs. (\ref{X'})-(\ref{L'}).

According to the center manifold theorem \cite{DynamicalReduction},
there exists a $C^\infty$-center manifold
\begin{eqnarray}
W^c_{loc}&=&\left\{(X,Y,Z,\bar{\lambda}): X\equiv h_1(Z,\bar{\lambda}), Y\equiv h_2(Z,\bar{\lambda}),\right.\nonumber\\
&&\left.h_i(0,0)=0, J_{h_i}(0,0)=0\right\}\nonumber
\end{eqnarray}
such that the dynamics of (\ref{A9}) can be restricted to the center
manifold. $J_{h_i}$ is the Jacobi matrix of $h_i$, and
$h_1(Z,\bar{\lambda}), h_2(Z,\bar{\lambda})$ are
\begin{eqnarray}
h_1(Z,\bar{\lambda})=A_1 Z^2+A_2 Z \bar{\lambda}+A_3 \bar{\lambda}^2+\cdots,\label{h1}\\
h_2(Z,\bar{\lambda})=B_1 Z^2+B_2 Z \bar{\lambda}+B_3
\bar{\lambda}^2+\cdots.\label{h2}
\end{eqnarray}
We just consider the quadratic forms of $Z$ and $\bar{\lambda}$ in
this appendix.

Considering the center manifold theorem, we have
\begin{eqnarray}
  \frac{dX}{dN}&=&\frac{\partial h_1(Z,\bar{\lambda})}{\partial Z}\frac{dZ}{dN}+\frac{\partial h_1(Z,\bar{\lambda})}{\partial\bar{\lambda}}\frac{d\bar{\lambda}}{dN},\label{X''}\\
  \frac{dY}{dN}&=&\frac{\partial h_2(Z,\bar{\lambda})}{\partial Z}\frac{dZ}{dN}+\frac{\partial h_2(Z,\bar{\lambda})}{\partial\bar{\lambda}}\frac{d\bar{\lambda}}{dN}.\label{Y''}
\end{eqnarray}
Inserting the Eqs. (\ref{h1}) and (\ref{h2}) into $dX/dN,dY/dN$ in
Eq. (\ref{A9}) and Eqs. (\ref{X''})-(\ref{Y''}), and comparing the
coefficients of $dX/dN$ and $dY/dN$, we get
\begin{eqnarray}
&&A_1=0,\quad A_2=\frac{\sqrt{6}}{6},\quad A_3=0, \quad B_1=\frac{1}{12},\nonumber\\
&& B_2=0, \quad B_3=\frac{1}{8}.
\end{eqnarray}
Then, the dynamics near the origin is governed by the following equations,
\begin{eqnarray}
\frac{dZ}{dN}&=&-{Z}^{3}f_1(0),\label{ZF'}\\
\frac{d\bar{\lambda}}{dN}&=&-{Z}^{2}\bar{\lambda}+\gamma\,{Z}^{2}\bar{\lambda}-\frac32\,\gamma\,{\bar{\lambda}}^{3}.\label{LF'}
\end{eqnarray}
We consider two different values of $\gamma$ to get the stability
properties of this system. This is because a different $\gamma$ will
give a different dynamical systems. The first one to be considered
is dust, which has $\gamma=1$. Then, we have
\begin{eqnarray}
   \frac{dZ}{dN}&=&-{Z}^{3}f_1(0)\label{ZF''},\\
   \frac{d\bar{\lambda}}{dN}&=&-\frac32\bar{\lambda}^3.\label{LF''}
\end{eqnarray}
According to Lyapunov's theorem, we can define a Lyapunov function
to analyze the stability properties of a dynamical system. Different
dynamical systems have different Lyapunov functions, and one
dynamical system can also have different Lyapunov functions. But all
the Lyapunov functions $U$ should satisfy $U(\mathbf{x})\geq 0$ at
the original point (Chapter 2 of \cite{Khalil-non}). Then we can
define
\begin{eqnarray}
  U_1=\frac12\left(Z^2+\bar{\lambda}^2\right).
\end{eqnarray}
Using Eqs.(\ref{ZF''}) and (\ref{LF''}), wehave
\begin{eqnarray}
 \frac{dU_1}{dN}=-f_1(0)Z^4-\frac32Z^2\bar{\lambda}^4.
\end{eqnarray}
According to Lyapunov¡¯s stability theorems, the system is stable if
$f_1(0)\geq 0$.

Now we turn to considering radiation, which has $\gamma=\frac43$.
Eqs. (\ref{ZF'}) and (\ref{LF'}) become
\begin{eqnarray}
\frac{dZ}{dN}&=&-{Z}^{3}f_1(0),\label{zc}\\
\frac{d\bar{\lambda}}{dN}&=&-2\bar{\lambda}^3+\frac13Z^2\bar{\lambda}.\label{lc}
\end{eqnarray}
We need to consider three possible cases: (a) $f_1(0)\neq 0$, (b)
$f_1(0)=0, Z(N=0)=0$, and (c) $f_1(0)=0, Z(N=0)\neq 0$, since these
three different cases will bring out three different dynamical
systems.

If $f_1(0)\neq 0$, the Lyapunov function can be defined as
\begin{eqnarray}
   U_2=\frac{1}{1+Z^2/(6A)+\bar{\lambda}^2},
\end{eqnarray}
where $A=f_1(0)$ if $f(0)>0$, and $A=-f_1(0)$ if $f_1(0)<0$. Then
one can get
\begin{eqnarray}
  \frac{dU_2}{dN}=\frac{12A^2\left[(Z^2-\bar{\lambda}^2)^2+5\bar{\lambda}^4 \right]}{\left[6A+6A\bar{\lambda}^2+Z^2\right]^2}>0.
\end{eqnarray}
Then this point is an unstable one.

If $f_1(0)=0$ and $Z(N=0)=0$, Eq. (\ref{lc}) becomes
$d\bar{\lambda}/{dN}=-2\bar{\lambda}^3$. Defining Lyapunov function,
\begin{eqnarray}
U_3=1+\bar{\lambda}^2,
\end{eqnarray}
then
\begin{eqnarray}
\frac{dU_3}{dN}=-4\bar{\lambda}^4\leq 0.
\end{eqnarray}

If $f_1(0)=0$ and $Z(N=0)\neq 0$, one can get $Z=C$ from Eq.
(\ref{zc}), with a non-zero constant $C$. Equation (\ref{lc})
becomes
\begin{eqnarray}
   \frac{d\bar{\lambda}}{dN}=-2\bar{\lambda}^3+\frac13C^2\bar{\lambda},
\end{eqnarray}
The Lyapunov function can be defined as
\begin{eqnarray}
  U_4=\left(1-\frac{6}{C^2}\bar{\lambda}^2\right)^2,
\end{eqnarray}
Then we have
\begin{eqnarray}
  \frac{dU_4}{dN}=-\frac{8}{C^4}\bar{\lambda}^2\left(C^2-6\bar{\lambda}^2\right)^2\leq 0.
\end{eqnarray}

Obviously, according to Lyapunov¡¯s stability theorem, this point is
stable as long as $f_1(0)=0$, regardless of $Z(N=0)= 0$ or
$Z(N=0)\neq 0$.

\end{document}